\documentclass[%
 aip,
 amsmath,amssymb,
preprint,
]{revtex4-1}

\usepackage{graphicx}
\usepackage{dcolumn}
\usepackage{bm}

\usepackage[utf8]{inputenc}
\usepackage[T1]{fontenc}
\usepackage{mathptmx}
\usepackage{etoolbox}

\makeatletter
\def\@email#1#2{%
 \endgroup
 \patchcmd{\titleblock@produce}
  {\frontmatter@RRAPformat}
  {\frontmatter@RRAPformat{\produce@RRAP{*#1\href{mailto:#2}{#2}}}\frontmatter@RRAPformat}
  {}{}
}%
\makeatother

\newcommand{\ie}{{\it i.e.}}
\newcommand{\eg}{{\it e.g.}}

\newcommand{\gsim}{\stackrel{>}{_\sim}}

\newcommand{\dd}{\mbox{\rm d}}

\newcommand{\eq}[1]{Eq.~(\ref{#1})}
\newcommand{\eqs}[1]{Eqs.~(\ref{#1})}
\newcommand{\noeq}[1]{~(\ref{#1})}
\newcommand{\fig}[1]{Fig.~\ref{#1}}
\newcommand{\figs}[1]{Figs.~\ref{#1}}
\newcommand{\nofig}[1]{~\ref{#1}}
\newcommand{\tabla}[1]{Table~\ref{#1}}
\newcommand{\sect}[1]{Sect.~\ref{#1}}
\newcommand{\sects}[1]{Sects.~\ref{#1}}
\newcommand{\nosect}[1]{~\ref{#1}}
\newcommand{\subsect}[1]{Subsect.~\ref{#1}}
\newcommand{\subsects}[1]{Subsects.~\ref{#1}}
\newcommand{\nosubsect}[1]{~\ref{#1}}

\newcommand{\beq}{\begin{equation}}
\newcommand{\eeq}{\end{equation}}

\newcommand{\EEE}{\mathcal{E}}

\usepackage{graphicx} 

\newcommand{\flujo}{\mbox{$\Phi$}}

\newcommand{\tref}{\mbox{$t_\mathrm{ref}$}}
\newcommand{\tsat}{\mbox{$t_\mathrm{sat}$}}

\newcommand{\Cimp}{\mbox{$C_{\rm imp}$}}
\newcommand{\Omegaclean}{\mbox{$\Omega_{0}$}}
\newcommand{\Omegacleanfinite}{\mbox{$\Omega'_{0}$}}
\newcommand{\nsat}{\mbox{$n_{\rm sat}$}}

\newcommand{\rhoeclean}{\mbox{$\rho_{\rm e0}$}}
\newcommand{\LRVinicial}{\mbox{LRV$_{0}$}}
\newcommand{\LRVinicialde}[1]{\mbox{LRV$_{0}^{\,\rm #1}$}}

\newcommand{\Dm}{\mbox{$\delta_{\rm m}$}}
\newcommand{\don}[1]{\mbox{\small($#1$)}}

\newcommand{\refsprimerafrase}{graule3,zirconia,JMS1,Tepper2,Tepper1,advances,Treccani,ACS2,ACS,Theron,mastepper,Toyota}

\begin{document}


\title{\mbox{}\vskip12pt\large Nanostructured micrometric-pore  membranes for nanofiltration: Micrometric geometry may optimize  performance,\\ energy efficiency and operational lifetime\vskip6pt\mbox{} }
\author{J.C. Verde}
\altaffiliation[In leave of absence]{}
\author{N. Cot\'on}
\altaffiliation[Currently at ]{Materials Physics Department, Complutense University of Madrid 28040, Spain}
\author{M.V. Ramallo}
\altaffiliation[Also at ]{Institute of Materials iMATUS, University of Santiago de Compostela 15782, Spain}
\email{mv.ramallo@usc.es}
\affiliation{QMatterPhotonics Research Group,   Department of Particle Physics, University of Santiago de Compostela 15782, Spain
}%
\homepage{http://qmatterphotonics.usc.es}


\begin{abstract}\noindent\setlength{\baselineskip16pt}\mbox{}\vskip6pt\mbox{}{\bf Abstract:}

Media and membranes composed of micrometric-diameter pores are well known in academia and industry to be capable of efficacious nanofiltration of fluids once the pore  inner surfaces are coated with nanostructures. Given the large mismatch between the two very different scales of these hybrid systems, it could be expected that trapping of nanoimpurities  would almost entirely depend on the characteristics of the nanostructures. However, we show here that the micrometric-scale nominal geometry does have noticeable impact on the nanofiltration performance, on  its evolution with time, and on the energy spent per trapped impurity. For that, we apply stochastic calculations customized to combine the cumulative probabilities of wall-impurity attraction and binding, supplemented with continuity equations as the fluid flows; this  allows tracking the nanofiltration without a many-particle simulation,  prohibitive in such multi-scale system. We focus on the influence of the micrometric nominal geometry over filtration features like  logarithmic removal value LRV,  operational lifetimes, energy balances, or spatial profiles of the trapped-impurity layer in the pore inner walls. Our results identify some  pore geometries (\eg, decreasing-conical and sinusoidal-corrugated) with about 4-fold larger  initial nanofiltering performance and operational lifetime than simple cylinders of the same average diameter. The  optimal geometry is also shown to depend on the LRV value acceptable for each specific application.

\end{abstract}

\maketitle

\setlength{\parskip}{6pt}

\section{Introduction}

One of the most innovative and promising techniques for water filtration is the application of nanotechnology.\cite{inicio1,inicio2,inicio3,inicio4,inicio5} Some examples of very interesting  results achieved by research in the topic include,  \eg, the fabrication of new nanoenhanced membrane and media,\cite{otrasmembranas1, otrasmembranas2,otrasmembranas3,otrasmembranas4,otrasmembranas5, Tepper1,Tepper2,graule3,mastepper,zirconia,advances,Treccani,ACS,ACS2,JMS1,Toyota,Theron} or the development of theory and  stochastic models\cite{modelos1,modelos2, modelos25,modelos3,modelos35,modelos4,electrostatics,EPL,GhosalA,GhosalD,GhosalB} customized to explain the behavior of novel filtrating elements and/or mechanisms.
For example, in recent years both experimentalists and industry\cite{\refsprimerafrase} have become progressively interested in membranes that achieve nanofiltration  by means of nanofunctionalization of the inner walls of pores with  diameters in the micrometric range $\sim$\,0.2--2\,$\mu$m. In these investigations\cite{\refsprimerafrase} it has been revealed the surprisingly good capability of these pores to filter out nanosized impurities from fluids, principally when the signs of the $\zeta$-potentials of impurities and functionalized pore walls are opposite so that  electrostatic effects enhance the filtration.\cite{\refsprimerafrase,electrostatics,EPL} This is in spite of such a relatively large diameter of the pores (note that for a pressure-driven flow passing through a cylindrical conduit with diameter 1\,$\mu$m, only about 0.04\% of the fluid will transit closer than 10\,nm from the walls). A key practical advantage of these novel nano/micro hybrid-scale membranes is that they are not affected by the very large hydrodynamic resistance of pores of nanometric-scale diameters. Also their saturation capacity  is much larger than for membranes of the same area but composed of nanometric-diameter pores. In fact, nanostructured micrometric pore membranes are currently commercialized for, \eg, energy-efficient nanofiltration of drinking water, clynical virus sampling, COVID masks of reduced breathing effort, industrial effluents treatment, etc.\cite{Theron,Tepper2,Tepper1,mastepper,Toyota}

It is currently possible to custom-engineer the geometry profiles of  micrometric-diameter pores of filtering membranes (see, \eg, Refs.~\onlinecite{Hamidreza,Xu,conical2,conical3,Wu} for a description of different ad-hoc geometries and the experimental techniques used to obtain them). These feasible designs include, among others, tubular conduits with cylindrical shapes having different diameters,\cite{Xu} corrugated tubes,\cite{Hamidreza} or pores with cone-like shapes.\cite{Wu,conical2,conical3} But to our knowledge such engineering has never been used with nanostructured micrometric pores. 

To our knowledge no experimental or theoretical investigation seems to exist up to now of the influence over the nanofiltration of the nominal micrometric   geometry of nanostructured micrometric pores, maybe because one could na\"{\i}vely expect that due to the large mismatch between the two very different scales of these hybrid systems, the nanofiltration characteristics  would almost entirely depend on the characteristics of the nanostructures, and not on the micrometric geometry.

In this article we present a theoretical study  of the influence of micrometric-scale nominal geometry in the nanofiltration characteristic of a single nanostructured micrometric pore, that reveals that micrometric geometry may in fact have considerable influence on various of the nanofiltration figures of merit, such as the logarithmic removal value LRV, the energy required for filtration, or the operational lifetimes. Our results should be directly applicable to membranes composed of such pores. They could be also useful for future investigations in three-dimensional porous media, where local tortuosity of the intensity and orientation of flow in each pore should be additionally considered. Also, investigating pores with different nominal geometries may also be relevant for random media, in which interconnections and throats between pores may be  studied in terms of conduits with sinus-like corrugated shapes.\cite{Balhoff,Maziar,Thauvin, Wangg}

For our study, we will apply stochastic calculations custom-adapted to combine cumulative probabilities of wall-impurity attraction and binding, supplemented with continuity equations as the fluid flows. For the wall-impurity attraction and binding probabilities, we employ  considerations adapted to the specific  attraction produced by  these electropositive (or negative) nanotextured walls over nanoimpurities of opposite sign, in the spirit of  the ones in  Ref.~\onlinecite{EPL} (for differential-length nanoenhanced micropores), the so-called lubrication model of Refs.~\onlinecite{GhosalA,GhosalD,GhosalB} (developed for electrophoresis applications), or other stochastic impurity-trapping models\cite{modelos4,modelos3,modelos35}. We supplement these probabilities with continuity equations for the flow and number of trapped and free impurities. Then we integrate these formulae in time and space employing the capabilities of modern parallel computers, what allows us to track the nanofiltration process without a many-particle molecular dynamics simulation (which would be much more  prohibitive  due to the multi-scale nature of the system). 

Our model calculates the evolution with position and time of the spatial profiles of the trapped-impurity layer in the pore inner walls, and the filtration figures of merit may be obtained from such evolution. Our results identify some  pore geometries (\eg, decreasing-conical and sinusoidal-corrugated) with about 4-fold larger initial nanofiltering performance and operational lifetime than simple cylinders of the same average diameter. The  optimal geometry is also shown to depend on the LRV value acceptable for each specific application. Some general trends seem to emerge from our simulations, like the tendencies for impurities to accumulate faster when closer to the entry point of the flow and  in the narrowest portions of the pores (these last two trends being either competing or cooperating depending on the specific considered geometry). The observed effect cannot be explained alone by the mere consideration of whether the fluid is made to flow closer to the nanostructures - it is instead a consequence of the special stochastics of the process (for instance, increasing and decreasing cones have different filtration evolution in spite of exposing the same nanostructured area and having the same percent of fluid flowing near a given distance from the walls).

\section{Models and Methods \label{methods}}
In this Section, we detail our methods to evaluate  the nanofiltration performance, the energy required for nanofiltration, and their  evolution with time, for different micrometric-scale nominal geometries. A rapid overview is as follows: We  build (\subsects{methods-initial} and~\nosubsect{methods-stochastics})  custom stochastic considerations that combine cumulative probabilities of electrostatic-type wall-impurity  attraction, and of binding once impurities collide with the nanostructures in the walls, taking into account the evolution with space and time  as flow passes and impurities progressively saturate the attractive and anchoring nanostructures in the wall. These stochastic considerations are then combined with continuity equations (\subsect{methods-continuity}), resulting in a model that allows to track the nanofiltration. We then write the relationships and methods (\subsects{methods-LRV} to~\nosubsect{methods-parameters}) to derive from these results the  relevant filtration properties such as  logarithmic removal value LRV,  operational lifetimes, energy balances, or spatial profiles of the trapped-impurity layer in the pore inner walls. 

\subsection{Initial considerations \label{methods-initial}}
Let us consider a tubular conduit of micrometric-scale diameter whose inner walls have a (initially homogeneous) nanotexture. The tube has a nominal diameter $d(x)$ that may, in general, vary along the axial coordinate $x$ (with $0\leq x\leq L$, where $L$ is the tube length). A fluid  passes  through, carrying  a certain concentration of nanoimpurities to be filtered out. All of the impurities are taken to be equal and have average radius $\rho_0$. Let us try  to obtain the evolution with space $x$ and time $t$ of the impurity concentration $\Cimp(x,t)$ in the fluid and of the areal density of impurities $n(x,t)$ trapped in the tubes (throughout this article, by ``areal density'' we refer to quantities normalized using the nominal area of the tube inner wall).

Initially ($t=0$) the system is in the ``clean state'', in which $n(x,0)=0$ and the diameter available to the flow is $d(x)$. As time and flow passes, impurities get trapped  and progressively cover the tube inner walls. Eventually, a ``saturated state'' is reached in which the nanotexture of the walls no longer electrostatically attracts further impurities. In the saturated state, $n$ is $n_{\rm sat}$ and the diameter available to the flow decreases down to a value $d(x_i)-\delta_{\rm sat}$.  We will consider in this article that after this saturated state the capability of the walls to anchor further impurities is zero (and hence the filtration performance)  so that we neglect conventional filtration mechanisms and focus only on the enhancement due to the  nanotexture.

In order to evaluate the effects of variations of the diameter along the tube's axial coordinate, and also to allow computation of successive spacial and temporal finite-difference iterations, we discretize the problem as follows: We divide the tube into $i=0,1,\dots N$ parallel slices of coordinates $x_i=(L/N)i$ and also discretize  time  as $t_{j+1}=t_j+\Delta t_j$ with $j=0,1,\dots$ (the reason to allow the time step to vary with $j$ will be to optimize our computational algorithms, see \sect{methods-computational}). 

\subsection{Stochastics \label{methods-stochastics}}

We will model the filtration process using the following stochastic description: We  consider a collision rate with the pore walls for the impurities flowing in the fluid, and also a binding probability to the nanotexture  once the impurity has collided with the wall, both quantities being time and space-dependent (\eg, via the number of impurities that remain nonsaturated by trapped impurities at each given instant and position). Let us consider both contributions in that order:

As mentioned in the introduction,  the good filtration rate of nanostructured micrometric pores cannot be explained if the only impurities colliding with the walls are those passing closer to them than  a distance of the order of the size of the impurities. Instead,   it is necessary  to consider the  effects of electrostatic attraction, that  enlarges the escape distance  below which the impurities will eventually collide with the nanotextured wall. We thus introduce a  collision distance  $\rho_{\rm e}$ as the  typical impurity-wall separation below which an impurity in the flow  will acquire  course of  collision with the wall. We expect $\rho_{\rm e}$  to grow with some form of effective charge  of the walls (proportional to their $\zeta$-potential) that will decay with time as impurities cover the wall and screen out the charges exposed in the nanotexturing. Using  simple electrostatic  escape distance arguments it may be expected  that $\rho_{\rm e}$ should   linearly grow with the effective charge of the walls. As shown, \eg, in Ref.~\onlinecite{EPL} (see also Ref.~\onlinecite{electrostatics}), a more precise law may be obtained when taking also into account the screening due to the Debye length of the fluid $\lambda_{\rm D}$:\cite{EPL,electrostatics} 
\begin{equation}
\rho_{\rm e}(x_{\rm i},t_{\rm j})= \rho_0 +\lambda_{\rm D} \,W\left( \frac{\rhoeclean-\rho_0}{\lambda_{\rm D}}   \left(1-\left\Vert\frac{n(x_{\rm i},t_{\rm j})}{n_{\rm sat}}\right\Vert \right) \exp\left(\frac{\rhoeclean-\rho_0}{\lambda_{\rm D}}\right) \right),
\label{eqrhoe}
\end{equation}
where $\rhoeclean$ is the collision distance in the clean state, $W(x)$ is the the principal Lambert function and the notation $\Vert x \Vert$ stands for  ${\rm min}(1, x)$. \cite{footnote-uno} Obtaining the ratio of impurities that travel within such  collision distance $\rho_{\rm e}$ from the walls is now a classical hydrodynamic problem, for which we shall use the typical Poiseuille profile of velocity distributions for moderate external pressure-driven flows. This leads to the fraction $f_{\rm e}$ of impurities that will collide the the $i$-th slice of the walls:
\begin{equation}
f_{\rm e}(x_{\rm i}, t_{\rm j})=\left( \left( \left\Vert \frac{ 2\rho_{\rm e}(x_{\rm i},t_{\rm j})}{d(x_{\rm i})-n(x_{\rm i},t_{\rm j})\delta_{\rm sat}/n_{\rm sat} }\right\Vert-1 \right)^2-1   \right)^2. 
\label{eqfe}
\end{equation}
In principle other more complicated velocity profile distributions could have been used but, as shown, \eg, in Refs.~\onlinecite{Poiseuille-1,Poiseuille-2}, the flow may be in fact expected to be in the Poiseuille regime if the radiuses of the pores remain $\gsim10$\,nm and the values of viscosity and   hydrostatic pressure are realistic for most filtering applications. (Note also that this is in contrast to the electrical-driven-like flow  assumed in the models   of Refs.~\onlinecite{GhosalA,GhosalD,GhosalB}  focused  in  electrophoresis scenarios). We are also implicitly assuming negligible turbulence and  axial advection on the transversal dynamics (as valid for moderate flow velocities; see also \subsect{methods-parameters} for a more comprehensive discussion of this latter aspect and the  implications over the parameter values we shall choose for our numerical evaluations). 

Concerning the impurity-wall binding probability for the impurities that do collide with the  walls, it has to start from a  maximum value  $\Omegaclean$ in the initial clean state and then  linearly decrease with the number of binding centers in the nanotextured walls, \ie, with $n(x,t)$.

Both of the aforementioned probabilities have to be multiplicatively accumulated so to obtain the fraction of impurities being trapped in each finite-element slice of the tube, at any given time and with the walls screened out by a concentration of previously trapped impurities. This results in
\begin{equation}
\label{fetrapped}
f_{\rm trapped}(x_{\rm i}, t_{\rm j})=\frac{L}{N}\,f_{\rm e}(x_{\rm i}, t_{\rm j})\,\Omegacleanfinite\left(1-\left\Vert\frac{n(x_i,t_i)}{\nsat}\right\Vert\right),
\end{equation}
where  
\begin{equation}
\Omegacleanfinite=(N/L)[1-(1-\rho_0 \Omegaclean)^{L/\rho_0 N}]
\label{eq-omegacleanfinite}
\end{equation}
is a correction to $\Omegaclean$ necessary when considering finite-element slices of length $\Delta x=L/N$.\cite{footnote-omegacleanfinite}

\subsection{Continuity equations and recursive finite-difference calculation of  $n(x_{\it i}, t_{\it j})$, $C_{\rm imp}(x_{\it i}, t_{\it j})$ and $\flujo(t_{j})$  \label{methods-continuity}}

By using the aforementioned probabilities, finite-difference-iterative  expressions may be now obtained for the areal density of trapped impurities $n(x_{\it i}, t_{\it j})$ in each $x_{\it i}$ slice of the tube at the time $t_{\it j}$, for a fluid flow rate $\flujo(t_{j})$ passing through the tube carrying  a concentration of  impurities $C_{\rm imp}(x_{\it i}, t_{\it j})$. Specifically, the continuity condition for the impurity number gives for the areal density of trapped impurities:
\begin{equation}
n(x_{i},t_{j+1})=n(x_{i},t_{j})+
\frac{\Delta t_j \Cimp(x_i,t_{j})\flujo(t_{j})\Omegacleanfinite}{\pi\, d(x_i)}
\left(1-\left\Vert\frac{n(x_i,t_{j})}{n_{\rm sat}}\right\Vert\right)  f_{\rm e}(x_{\rm i}, t_{\rm j}).
\label{diferencialred}
\end{equation} 
and for  the concentration of impurities in the fluid arriving to the next $\Delta x$ slice:
\begin{equation}
\Cimp(x_{i+1},t_{j})=\Cimp(x_{i},t_{j})-\frac{\pi  L\; d(x_{i})}{N\Delta t_{j}\flujo(t_{j})} \Big(n(x_{i},t_{j})-n(x_{i},t_{j-1})\Big).
\label{continuidad}
\end{equation}

For the fluid flow rate $\flujo(t_j)$, as already mentioned above we will assume in this article that the liquid is driven by a constant hydrostatic pressure difference $P$ between opposite ends of the tubular conduits. Thus, we use a Poiseuille relationship:\cite{EPL,Poiseuille-1,GhosalA}
\begin{equation}
\flujo^{-1}(t_j)=\frac{128\eta L}{\pi PN}\sum_i\left(d(x_i)-\frac{n(x_i,t_j)\,\delta_{\rm sat}}{\nsat}  \right)^{-4}.
\label{flujo}
\end{equation}
Note that the flow rate becomes $t$-dependent even with constant $P$, due to the progressive narrowing of the pore as it becomes dirtier. Eventually, if $d(x_i)-n(x_i,t_j)\delta_{\rm sat}/n_{\rm sat}=0$ at some $x_i$, the conduit would clog and $\flujo=0$.

\subsection{Figures of merit for filtration functionality: Logarithmic removal value LRV, operational lifetimes and energy consumption per trapped impurity \label{methods-LRV}} 

As customary,\cite{Treccani,JMS1,ACS,ACS2,Tepper1,Tepper2,graule3,zirconia,mastepper,advances,Toyota} we will characterize the filtration performance through the so-called logarithmic removal value:
\begin{equation} \label{LRVdef}
{\rm LRV}(t)=-{\rm log}_{10}\frac{\,\Cimp\don{x=L,t}}{\,\Cimp\don{x=0,t}}.
\end{equation}
For instance, 99.9\% impurity removal corresponds to LRV=3. We will use the notation \LRVinicial\ for  LRV($t=0$), corresponding to the clean state of the filter. It is interesting to note that it is possible  within our stochastic model to obtain explicit expressions for \LRVinicial\ for each of the geometries considered in this article. We describe them in an Appendix. We have checked that application of those direct equations provide  the same results as the $t\rightarrow0$ limit of our numerical finite-difference evaluations presented in Sect.~\ref{results}.

To characterize the endurance of the filter, let us  introduce the  LRV$\geq$5, LRV$\geq$2 and LRV$\geq$1 lifetimes,  defined as the $t$-values at which the LRV becomes 5-log, 2-log and 1-log, respectively. In practice, which of these  endurance criteria (5-log to 1-log) is more relevant will depend on the specific  application intended for the filter (\eg,  pathogen removal down  to clinically safe levels vs.~pre-filtration stage in industrial effluent treatments). Also, we  will find useful to define the times at which certain thresholds are reached for the concentration of trapped impurities in the pore walls. So, we introduce the times $t_{0.15}$, $t_{1/2}$ and 
$t_{\rm sat}$ as those when  the average areal density of trapped impurities in the whole tube, $\bar{n}(t)$, becomes respectively 0.15$n_{\rm sat}$, 0.5$n_{\rm sat}$ and $n_{\rm sat}$. 
 
 One of the key advantages of nanostructured micrometric pore filters vs.~alternatives such as size-exclusion nanopore membranes is their lower hydrodynamic resistance and correspondingly lower pressure required to achieve the fluid flow. Therefore, a quantity of importance is the energy consumed by the filtration, $E$, that can be evaluated by integrating in time the product of the flow rate and external pressure (\ie, the hydrodynamic dissipated power). \cite{footnote-dos} Probably still more important  is the energy normalized by the number of trapped impurities, henceforth noted as $\EEE$. It is not difficult to arrive to the following relationship linking the time evolutions of $\EEE$ and LRV:
\begin{equation}\label{hydrod}
\EEE(t)=\frac{P}{      C_0    \left(      1-10^{-{\rm LRV}(t)}       \right)           }.
\end{equation}

\subsection{Computational procedures  \label{methods-computational}}
The \eqs{diferencialred} to\noeq{flujo} form a set that may be   computationally integrated iteratively to obtain the $x$- and $t$-evolution  of the   pore filter. For that purpose, we implemented  ad-hoc software that accepts arbitrary user-defined  geometries for the nominal diameters $d(x)$. It iterates in both the $j$ and $i$ indexes (parallelizing the process so to use GPU accelerators), continuously calculating self-adaptively the time steps $\Delta t_j$  by requiring that successive instants vary $n(x_{\rm i},t_{\rm j})$ less than 0.01\% (for all $x_{\rm i}$ values).  This results in about $10^4$ time-intervals for each simulation. For the spacial discretization, we divided the tubular pores into $N=10^6$ finite-difference slices (this leads in our simulations to a trapping probability for each  impurity colliding with the walls always below $10^{-4}$ per finite-difference $x$-slice). 

For the boundary conditions of our simulations we take as initial state $n\don{x,0}=0$ (\ie, clean state for $t=0$) and $\Cimp\don{x\neq0,0}=0$. For the spatial boundary condition we take $\Cimp\don{0,t}=C_0$ ({\mbox{\it i.e.}}, a constant impurity concentration for the fluid at the entry of the pore).

\subsection{Typical parameter values \label{methods-parameters}} 

For the various parameters describing the  characteristics of the nanofunctionalized micrometric pore conduits in \eqs{diferencialred} to\noeq{flujo} we  use values of the same order of magnitude as the ones that were used in Ref.~\onlinecite{EPL} to discuss (but not considering  the effects  of the pore micrometric geometry) available systems\cite{graule3,zirconia,Tepper1,Tepper2,mastepper,Toyota} with  nanocoatings of varied compositions. We  use  nanoimpurity radius $\rho_0=10$~nm,  effective collision distance in the clean state $\rhoeclean=30$\,nm,  binding probability in the clean state $\Omegaclean=10^{-4}/$\,nm,  saturation impurity layer thickness  $\delta_{\rm sat}=40$\,nm and  saturation impurity areal density  $\nsat=10^{-2}\,\mbox{nm}^2$. For the  Debye length, we will use $\lambda_{\rm D}=10$\,nm. 

We also use an incoming impurity concentration  $C_0=10^{10}/\mbox{m}^3$,  the viscosity typical  of water $\eta=10^{-3}$\,Pa\,s, and a pressure difference  $P=10^5$\,Pa,  values that are  of the same order as the  experimental conditions imposed in Refs.~\onlinecite{graule3,zirconia,Tepper1,Tepper2,mastepper,Toyota}.  But we have checked that the results of our simulations are invariant with respect to these last three parameters except for a multiplicative rescaling of the time values. This effect is due to the symmetry of the starting equations. Consequently, in the present article we shall preferentially express  the times resulting from all our simulations normalized with respect to a common reference value  $t_{\rm ref}$. In particular, for this $t_{\rm ref}$ we will choose, for all the considered geometries (and as described in detail in \sect{cilindros}), the half-saturation time $t_{1/2}$ of the cylindrical cylinder with diameter 300\,nm. Increases in $\eta$ would  increase  $t_{\rm ref}$ linearly, while increases in $C_0$ and $P$ would  induce an inversely proportional change in  $t_{\rm ref}$.

Concerning the size of the pores, we will always use a millimetric length $L=1$\,mm. For their diameter, in general we will use   minimum, average and maximum values of, respectively, 200\,nm, 300\,nm and 400\,nm. In the case of the diameter being constant (\ie, cylinder pores) we will also check the effect of increasing it, scanning from 300\,nm to 400\,nm.  (Note  here that, with these  pore dimensions, the area covered by anchoring centers, \ie, the inner surface of the pore, is about four orders of magnitude larger than the transversal area; this alone  explains already why the saturation capacity of a membrane of nanostructured micrometric pores is well larger than the one   of a equivalently-sized membrane  composed of nanometric pores filtering  by transversal size-exclusion effect.)

Finally, let us briefly comment at this point  that the above typical parameter values (especially  pressure) correspond  to moderate flow velocities, in particular small enough for the assumption  that axial advection effects do not complicate the recovery of the equilibrium in the transversal  concentration profile  of impurities as they are filtered out.\cite{EPL} This is because the typical time it takes for diffusion to rebuild transversal equilibrium is orders of magnitude smaller than the typical time each impurity travels within the collision distance from the walls before being removed from the liquid: The first of those times may be estimated as $\lambda^2/D$,\cite{difussion} where $D$ is the diffusion coefficient ($\sim 10^{-9} {\rm m}^2$\,s) and $\lambda$ the typical transversal distance over which equilibrium has to be reestablished (that may be taken somewhere between $\lambda_{\rm D}$ and $\rho_{\rm e}$, \ie, 10--30\,nm). This leads to times of the order of $10^{-6}$\,s. In contrast, the typical time that each impurity travels within the collision distance from the walls before being removed from the liquid may be estimated to be about $4\times10^{-2}$\,s (by  combining an average Poiseuille velocity $Pd^2/(32L\eta)$, a typical diameter $300$\,nm and a typical distance traveled before being trapped $ \Omegaclean^{-1} =  10^{-5}$\,m).


\section{Results and discussion} \label{results}
\subsection{Cylindrical conduits as a function of diameter\label{cilindros} \label{section3.1}}

We first explore the filtration characteristics of  cylinders as a function of their diameter.  For that purpose, we computationally integrated our equations using for the diameter function $d(x)$  different  constant values  (the rest of the  parameters being, for all the cylinders, equal to those detailed in \sect{methods-parameters}).   The \tabla{tablauno} and the panels (a) to (d) of  \fig{figurauno} display the results of our simulations for representative example diameters. In particular, the left-hand column  (\ie, panels (a) and $c$) shows the internal profile of the trapped-impurity layer in each tube at different instants of the time evolution, \ie, their true internal radius as the conduits become narrower due to the accumulation of trapped impurities. These profiles are shown for the following instants of time: For $t=0$ or clean state (when the  radius  coincides with its nominal value $d(x)/2$);  for $t=t_{\rm sat}$, {\mbox{\it i.e.}}, when all of  the conduit has reached its saturated state (and the internal radius equals $(d(x)-\delta_{\rm sat})/2$); and also for the two intermediate times $t_{0.15}$ and   $t_{1/2}$ (when   the average areal density of trapped impurities in the whole tube $\bar{n}(t)$ equals 0.15$n_{\rm sat}$ and 0.5$n_{\rm sat}$). 

A first result of our simulations,  already easily visible in the panels (a) and $c$ of \fig{figurauno}, is that the impurities accumulate first at the beginning of the cylinders, instead of uniformly.

\begin{table}[b]
\caption{\label{tablauno}\small\setlength{\baselineskip}{16pt} Logarithmic removal value  at the initial $t=0$ clean state \LRVinicial, and operational lifetimes for different LRV thresholds,  for nanofunctionalized micrometric pores of different geometries and average diameters.  The times are expressed normalized by the same  common reference  \tref\  as  in \figs{figurauno} and\nofig{figurados}.\vspace{3pt}}

\begin{ruledtabular}
\begin{tabular}{ccccc}
pore geometry                       &                       & LRV$\geq5$             & LRV$\geq2$         & LRV$\geq1$ 
\\
\& average diameter    &  \LRVinicial & lifetime                      & lifetime                  & lifetime  \\
\hline
cylinder    300\,nm     & 5.6                & 0.17 $t_{\rm ref}$   & 1.28 $t_{\rm ref}$ & 1.71 $t_{\rm ref}$ \\
cylinder    350\,nm     & 4.3                & -                                     & 0.77 $t_{\rm ref}$   & 0.95 $t_{\rm ref}$ \\
cylinder   400\,nm     & 3.3                & -                                     & 0.39 $t_{\rm ref}$   & 0.53 $t_{\rm ref}$  \\
\hline
increasing  cone 300\,nm&  6.2 	    &  0.22 $t_{\rm ref}$   &    1.79  $t_{\rm ref}$      & 2.93 $t_{\rm ref}$\\
decreasing  cone 300\,nm&  6.2 	    & 0.76 $t_{\rm ref}$   &    2.18  $t_{\rm ref}$      &   2.53 $t_{\rm ref}$\\
sinusoid-corrugated 300\,nm&  6.4 & 0.64 $t_{\rm ref}$   &    2.66  $t_{\rm ref}$      &   3.56 $t_{\rm ref}$
\end{tabular}
\end{ruledtabular}
\end{table}

The right-hand column of \fig{figurauno}  shows the time-evolution of the areal density of trapped impurities at opposite ends of the cylindrical tube, $n(x = 0,t)$ and $n(x = L,t)$. It also shows the average areal density $\bar{n}(t)$ used to calculate the times $t_{0.15}$ and   $t_{1/2}$ (that are marked as solid points in that curve). The timescale in this figure is normalized as $t/t_{\rm ref}$, where $t_{\rm ref}$ is chosen as the $t_{1/2}$ value of the cylinder with diameter 300\,nm; the same common $t_{\rm ref}$ will be used in all of the figures and tables of the present work.

These representations evidence that for cylindrical conduits the growth with time of $n(x,t)$ happens earlier when closer to the start of the tube $(x = 0)$ than when closer to the exit point $(x = L)$, by at least one order of magnitude in time value (note that the time axis is logarithmic). As it could be expected, the behavior of $\bar{n}(t)$ is intermediate between the ones at both edges of the conduit.

The results of our simulations also serve to study the influence of the diameter of the cylinders over the filtration performance and operational lifetimes given by the logarithmic removal value LRV. In \tabla{tablauno} we list for each diameter the LRV value computed at the initial time, \LRVinicial, that   characterizes the maximum filtration capability of the cylinder (achieved in the initial, clean  state). It may be noted that smaller diameters lead to  better \LRVinicial. The \tabla{tablauno}  also lists the times at which  the LRV reaches the thresholds LRV$=5$, LRV$=2$ and LRV$=1$.  Logically, the LRV $\geq 5$ lifetime is null for the cases in which $\LRVinicial < 5$. Our results indicate  that these lifetimes become significantly larger when the diameter becomes smaller, in line with the trend of \LRVinicial. This also makes sense when comparing to each other the panels (b) and (d) of \fig{figurauno}: this comparison evidences a similar type of change in the times at which the $n(t)$ curves reach the saturation value (and, in fact, $t_{0.15}$ and $t_{1/2}$). 

The time evolution of the LRV performance is represented in further detail in \fig{figurados} for the cylinder with 300\,nm diameter. 

The evolution  of the energy $\EEE$ consumed per trapped impurity as a function of the amount of  impurities that got accumulated into the pore is represented in \fig{figuratres} for the cylinder with 300\,nm diameter. It may be noticed that it only increases considerably past the LRV$=2$ lifetime (a result that in fact will hold for all the geometries studied in this work). The initial, clean-state value $\EEE(t=0)$ is almost the same (well within 0.1\%) for all the studied cylinder diameters.

\begin{figure*}
\begin{center}\includegraphics{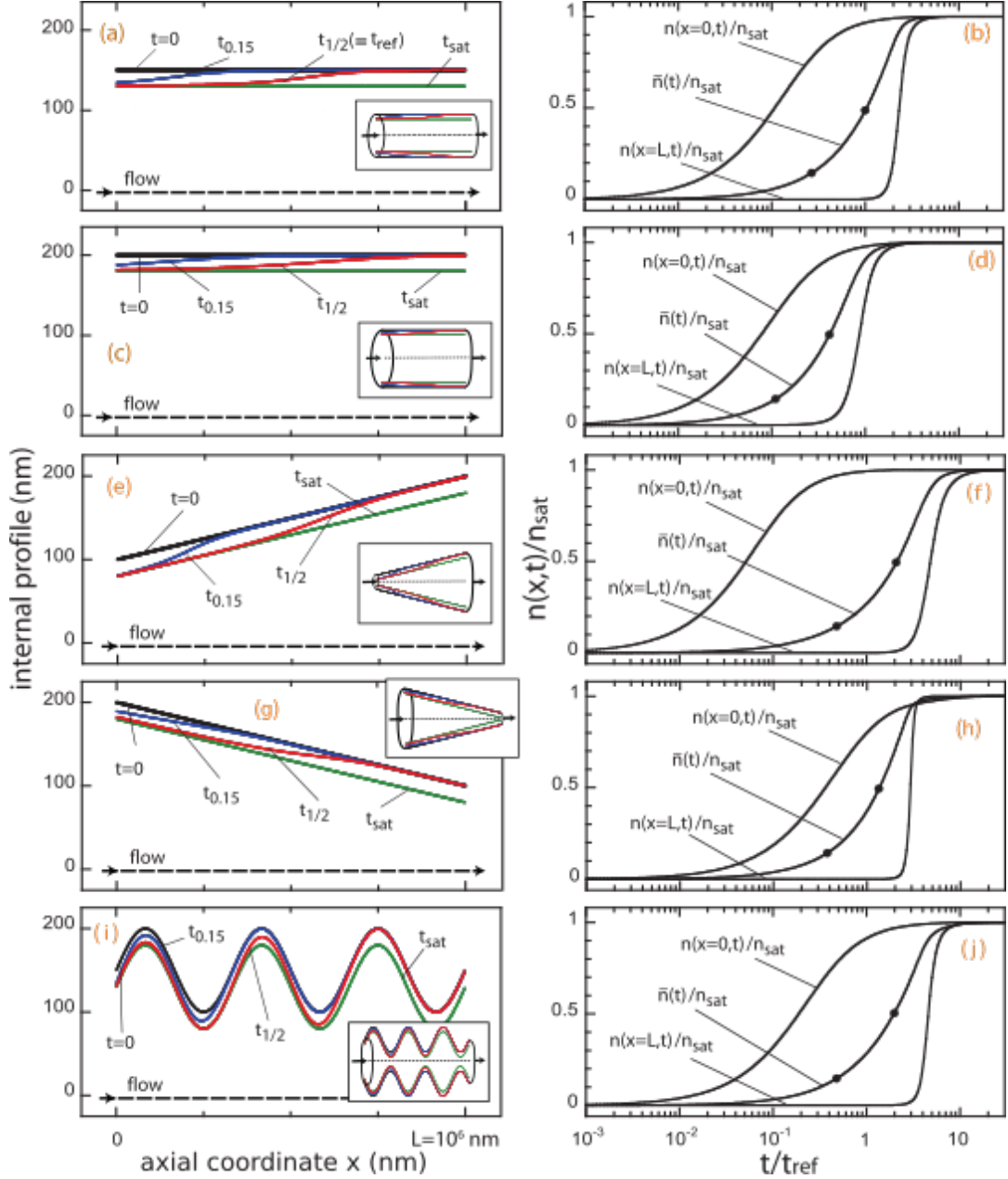}\end{center}
\caption{\label{figurauno}\small\setlength{\baselineskip}{16pt} Evolution of the layer of trapped impurities as flow passes through the nanostructured  pores of  the micrometric geometries indicated in the insets (cylinders of  diameters 300 and~400\,nm, increasing and decreasing cones, and sinusoidal corrugations). Left column: spatial profiles of the adsorpted layer, for selected  times. Right column:  time evolution of the areal density $n$ of trapped impurities, for the entry and  exit points  and for the spatial average $\bar{n}$. The times $t_{0.15}$ and $t_{1/2}$ are defined by $\bar{n}=0.15n_{\rm sat}$ and $0.5n_{\rm sat}$ (marked as solid points in the second column) and \tsat\ corresponds to saturation ($\bar{n}=\nsat$). The time axis is normalized by a common   \tref\ defined as $t_{1/2}$ of the 300\,nm cylinder. Note the (sometimes competing) tendencies of accumulation being faster both near the entry of the pore and in its narrowest points.}
\end{figure*}

\begin{figure}
\begin{center}\includegraphics[]{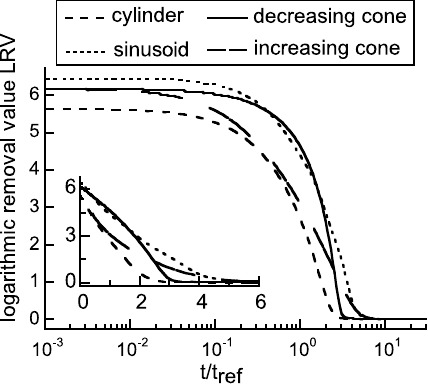}\end{center}
\caption{\label{figurados}\small\setlength{\baselineskip}{16pt}Logarithmic removal value LRV as a function of time obtained for  nanofunctionalized micrometric pores with different geometries and average nominal diameter 300\,nm. Note that the cylinder leads to the lowest initial value, \LRVinicial, and  lowest operational lifetimes (see also \tabla{tablauno}). The time axis is normalized by the same common reference  \tref\ as in \figs{figurauno} and \tabla{tablauno}.}
\end{figure}

\subsection{Cones of diameter increasing in the direction of flow.\label{conos-crecientes}\label{section3.2}}
Let us now show that other conduit geometries may improve the filtration performance and/or operational lifetimes with respect to those obtained for cylinders. We begin that investigation by studying  conical tubular conduits with diameter increasing in the direction of the fluid flow. Note that these are indeed feasible micrometric geometries: there exist  experimental realizations of membranes with conical (increasing and decreasing) pore shapes~\cite{Wu,conical2,conical3} for different applications, such as the study of resistive-pulse biosensors\cite{conical2} or  capillary electrophoresis\cite{Wu}. However, to our knowledge there are not yet investigations of the  nanofunctionalization of the inner surface of conical-shape pores  for  filtration applications.

In panels (e) and (f) of \fig{figurauno} we present the results obtained by performing our simulations with the $d(x)$ function corresponding to the increasing-conical  geometry. In particular, the diameter increases linearly from the value 200\,nm at the entrance to the tube up to the value 400\,nm at the exit point (this is to be compared therefore with the 300\,nm cylinder). Again it is visible that  impurities accumulate sooner at the entrance of the tubes than at their exit. But it can be noticed in the figures that the differences are now even larger than in the cylindrical case.  As a consequence, the $\bar{n}(t)$ evolution also spreads over a wider $t$-range.

As it may be noticed from the values listed in \tabla{tablauno}, the initial filtration performance \LRVinicial\  achieved by the increasing cone improves the one of  the equivalent  300\,nm cylinder (6.2-log versus 5.6-log, or about 4 times smaller concentration of impurities in the fluid at the exit of the pore).

\begin{figure}[t!]
\begin{center}\includegraphics[]{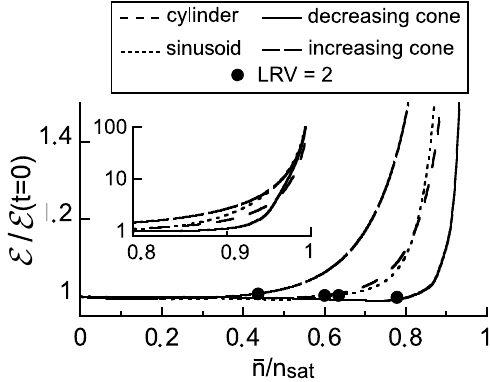}\end{center}
\caption{\label{figuratres}\small\setlength{\baselineskip}{16pt}Energy per filtered impurity obtained for  nanofunctionalized micrometric pores with different geometries and  average nominal diameter 300\,nm, normalized to its initial  value in the same pore (or in the reference 300\,nm cylinder; the difference would be unobservable in the figure). Note the  increase after the LRV$=2$ threshold (solid dots), that in turn is reached (see \tabla{tablauno}) at quite different times for each geometry. }
\end{figure}

The \tabla{tablauno} also indicates that the increasing-cone geometry produces  operating lifetimes larger by about 30\% (for the LRV$\geq5$ criterion), 40\% (for the LRV$\geq2$ criterion) or 70\% (for the LRV$\geq1$ criterion) with respect to  cylinders with the same average diameter. These differences are also illustrated in \fig{figuratres}, that compares the LRV$(t)$ curves of conduits with different shapes (but with equal average diameter 300\,nm). We note again that the time axis in all our figures is expressed normalized by a common reference value (\tref\ defined as  $t_{1/2}$ of the $d=300$\,nm cylinder) and in logarithmic scale.

These results indicate, therefore, that a possible path to greatly increase the operational lifetime of nanoenhanced  pore filters is to employ  supporting  micrometric geometries with variable  diameter. In the following Sections this conclusion will be confirmed with other  pore shapes.

In relation to the energy consumption per filtered impurity, \fig{figurados} shows that (as already found for cylinders) it only increases appreciably as time passes the LRV=2 lifetime. Also the initial value of $\EEE$ is similar to the  cylinder with the same average diameter (would be indistinguishable in the horizontal axis range of  \fig{figurados}, as in fact will happen for all geometries studied in this work; for that reason the figure just expresses $\EEE$ normalized to its initial value). Therefore, the main difference with respect to the equivalent cylinder is that such  LRV=2 threshold is reached, as mentioned before, about 40\% later in time.


\subsection{Cones of diameter decreasing in the direction of flow.\label{conos-decrecientes}\label{section3.3}}

We  consider now a decreasing-conical geometry, \ie, a linearly decreasing diameter $d(x)$ function. Again we use in our simulations 200 and 400\,nm for the extreme diameter values, and the mean diameter along the axial direction remains 300\,nm.  Panels (g) and (h) of \fig{figurauno} illustrate the results of our simulations for that geometry. A significant qualitative difference with respect to the previous increasing-cone conduit is that now the plots versus time of $n(x=0,t)$ and  $n(x=L,t)$ are significantly closer to each other. In fact, near the saturation state both lines even cross each other. This means that no longer is always true that the beginning of the conduit, $x=0$, accumulates impurities faster than its end, $x=L$. Instead, the smaller section at the end of the cone is able to overcome the rhythm of impurity accumulation of the first slices of the cone before these get saturated.

This more uniform filling by impurities results in a different evolution with time of the filtration performance. As revealed by the values listed in \tabla{tablauno} (see also \fig{figuratres}), the initial LRV values of the increasing and decreasing cones are coincident, but the LRV$\geq5$ lifetime of the decreasing cone is even longer than the one of the increasing cone (about 3.5~times longer than in the increasing cone, and about 4.5 times longer than in the cylinder of equal mean diameter). In contrast, the LRV$\geq1$ lifetime of decreasing cones is shorter that the one of increasing cones (about 15\% shorter, though still about 50\% longer than for equivalent cylinders). On the other hand, the LRV$\geq 2$ criterion reveals an intermediate tendency, getting a higher lifetime in comparison both with the cylinder (70\% higher) and the increasing-cone (20\% higher). This can be understood by observing  in \fig{figuratres} that  the LRV$(t)$ curves of both types of conical shapes cross each other near LRV$\sim1.5$. 

In relation to the  energy consumed per impurity filtered out, this geometry follows some trends similar as cylinders and increasing cones: The initial value is similar to those geometries and again (see \fig{figurados}) it almost does not increase with time until the  LRV=2 point. However, it also presents some differences in that such LRV=2 point happens at considerably larger $\overline{n}/n_{\rm sat}$ and that beyond this point decreasing cones actually degrade faster than cylinders and increasing cones.


\subsection{Sinusoid-corrugated  tubular conduits.\label{sinusoidales}\label{section3.4}}

Let us now consider corrugated tubular-like  pores, of diameter increasing and decreasing successively along their axis. Again this micrometric pore geometry is experimentally realizable and studies do exist on  networks composed by (non-nanostructured) micrometric pores with spherical shape  and interconnections (throats) of small radius forming sinus-like geometries.\cite{Hamidreza} Also, some theoretical studies have analyzed  porous systems as a network of pores with sinus-like shapes.\cite{Balhoff, Maziar}

A $\sin(x)$-like diameter functionality is possibly one of the simplest of the corrugated geometries, and could a priory be expected to serve as a fair first approximation to, at least, the most essential features of real corrugated pores.\cite{Hamidreza, Balhoff, Maziar} In panels (i) and (j) of  \fig{figurauno} we present the results for our simulations using for the diameter profile the function $d(x)=(300\,{\rm nm})+(100\,{\rm nm})\sin(6\pi x/L)$, \ie, the maximum, minimum and  mean diameter are the same as in the previous two Subsections, and $d(x)$ presents three oscillations over the conduit's length (we have checked that increasing the number of oscillations would not  qualitatively affect our results; note also that according to \eq{LRVsinus} \LRVinicialde{sinus} does not depend on the number of oscillations).

The time-evolution of the internal profiles shown in panel (i) of \fig{figurauno} evidences that in the corrugated conduits  the impurities accumulate first in the initial oscillation periods, and within that tendency they accumulate first in their narrower portions.  This result is coherent with a na\"{\i}ve vision of crudely considering the corrugated conduits as a combination in series of alternately increasing and decreasing cones. The time-evolution of the average $\bar{n}(t)$ and of the filtration performance LRV$(t)$ also resembles somewhat the results obtained for cones. However, it must be noted that the initial LRV performance is better than the one of the cylindrical and conical shapes (see \tabla{tablauno} and \fig{figuratres}). The difference is almost of one logarithmic unit with respect to the cylinder. Concerning the operational lifetime, with the LRV$\geq5$ criterion it is almost as long as in the decreasing cone (16\% shorter) and still much longer than for cylinders (almost 4 times longer). This  is  achieved without penalizing the LRV$\geq1$ lifetime: on the contrary, with the LRV$\geq1$ criterion the operational lifetime is even longer (about 20\%) than for decreasing cones. The LRV$\geq2$ lifetime  is also favorable (about double than for 300\,nm cylinders).

In terms of the hydrodynamic energy rate, it can be seen in \fig{figurados} that for $\overline{n}/n_{\rm sat}< 0.9$ the sinusoid geometry mimics quite well the behavior of its equivalent cylinder. However, for higher values, the sinusoid becomes energetically less efficient, and starts to become more similar to the increasing cone, which is the one less efficient. Of course, other difference with the cylinder is that the LRV=2 mark is reached at significantly later time, as already mentioned.

\section{Conclusions}\label{section4}

To sum up, we have studied the nanofiltration achieved  by individual  pores   of micrometric diameter with nanostructured inner walls, as a function of the micrometric-scale geometry. We focused on their logarithmic retention value LRV in the initial or clean state, on its evolution with time, on their operational lifetime for different acceptable LRV thresholds, on the energy consumed by the filtration, and on the evolution of the inner radius of the pores as impurities accumulate over their  walls. For our studies, we wrote stochastic equations taking into account the special attraction effect of the nanotextured surfaces due to their electropositivity (or negativity) and also the necessary continuity equations for the flow and trapping of nanoimpurities. We  considered different geometries, including first cylinders of different diameters and then pores with diameter varying along the direction of the fluid flow (increasing and decreasing conical tubes and sinusoid-corrugated tubes). 

Our results indicate that, in spite of the large mismatch between the nano- and micro-scales of these systems, the nanofiltration performance and its time-evolution do depend also on the micrometric geometry. 

With regard to the diameter size for cylindrical micrometric pores, we found that decreasing it improves  \LRVinicial\ (and this  enlarges the operational times during which the LRV is above a  threshold.)

Concerning pores with  micrometric diameter varying along the flow direction, we found that these variations allow to improve the filtration characteristics without decreasing the average diameter of the conduit. In particular {\it i)}~increasing-conical conduits have, with respect to  cylinders with the same average diameter,   better  initial LRV and significantly longer operational times (about 30-70\% longer, depending on the acceptable LRV threshold); {\it ii)}~decreasing-conical conduits equally improve the initial LRV and extend even further the operational times for high-LRV filtration requirements (more than  4.5 times longer than equivalent cylinders) while the low-LRV operational lifetime increases but to a lesser extent; and {\it iii)}~sinus-corrugated tubular conduits improve the initial LRV values by almost one logarithmic unit with respect to equivalent cylinders, and   extend the operational lifetimes for both  high- and low-LRV requirements (almost 4 times and about 2 times longer than equivalent cylinders, respectively). All these geometries also extend, with respect to cylinders of equal average diameter, the period of time in which energy consumption per trapped impurity is favorable.

Our results  also reveal  two general tendencies in the process of impurity accumulation in the walls, that may be synergic or antagonist with each other depending on the specific geometry considered: The first trend is that  the nanoenhanced walls get covered by nanoimpurities at a faster pace in the regions with smaller diameters; the second is that they get covered faster when nearer to the entrance of the flow in the pore. When both types of regions do not coincide in space, their competition may significantly increase some operational lifetimes. For example, when comparing  increasing and decreasing conical pores, in spite of having equal initial LRV, their time evolution is significantly different from each other (note, \eg, their LRV$\geq5$ lifetimes  in \tabla{tablauno}).

Therefore, our results indicate that engineering of the micrometric gometry of pores (as in fact, experimentally achieved by several groups with pore shapes similar to the ones studied in this work\cite{Wu, conical3, conical2, Hamidreza, Xu})  should be applied to nanostructured micrometric pore filters to optimize their nanofiltration characteristics. This is specially so in the case of membrane filters, where our results should be directly applicable (whereas for 3D media it may be expected that the specific  tortuosity and flow locality of each given material and morphology should be additionally taken into consideration).

\begin{acknowledgments}
We acknowledge support by University of Santiago de Compostela, Project 2024-PU036 `Propiedades de materiais supercondutores micro e nanoestruturados'.  
\end{acknowledgments}

\mbox{}

\mbox{}

\section*{\it Author Contributions}
{\footnotesize\noindent
{\em J.C. Verde:} Conceptualization (equal); Formal analysis (equal); Methodology (equal); Software (equal); Visualization (equal); Writing -- original draft  (equal). {\em N. Cot\'on:} Conceptualization (equal); Formal analysis (equal); Methodology (equal); Software (equal); Visualization (equal); Writing -- original draft  (equal). {\em M.V. Ramallo:} Conceptualization (lead); Formal analysis (equal); Methodology (equal); Software (equal); Visualization (equal); Writing -- original draft  (equal); Writing -- review \& editing (lead).}

\newpage

\appendix* 

\section{\label{apendice} Analytical expressions for \LRVinicial}
Explicit analytical expressions for \LRVinicial\  can be calculated for each of the geometries studied in \sect{methods}.  For that,  multiply over all $x$-positions the survival no-trapping probabilities to first obtain:
\begin{equation}
{\rm LRV}(t) = -\sum_{i}\log_{10}\left[1-\Omegacleanfinite \left(1- \left\Vert \frac{n(x_{\rm i},t_{\rm j})}{n_{\rm sat}} \right\Vert \right) 
\left[ \left(\left\Vert \frac{ 2 \rho_{\rm e}(x_{\rm i},t_{\rm j})}{d(x_{\rm i})-n(x_{\rm i},t_{\rm j})\delta_{\rm sat}/n_{\rm sat} }\right\Vert-1 \right)^2-1   \right]^2 \Delta x \right].
\end{equation}
For the initial, clean state ($t=0$, $n(x_i,0)=0$ and $\rho_{\it e}(x_i,0)=\rhoeclean$) this reduces to:
\begin{equation}\label{LRVgen}
\LRVinicial= (\ln10)^{-1} {\displaystyle\int_{x=0}^{L}} \Omegaclean \left[ \left( \left\Vert \frac{2 \rhoeclean}{d(x)}\right\Vert-1 \right)^2 -1   \right]^2 \mbox{\rm d}x ,
\end{equation}
where we also recovered the continuous  model by using $\Delta x\rightarrow\dd x$ and the power expansion $\log_{10}(1-y\mbox{\rm d}x)\approx -y\mbox{\rm d}x/\ln10$. Let us now carry out his integration for each given geometry.

For  cylindrical pores of diameter $d= \rhoeclean \delta $ (the $\rhoeclean$ normalization achieves compactness here):
\begin{equation}\label{LRVcyl}
\LRVinicialde{cylinder} =\frac{16 \Omegaclean L }{\ln{10}}\left[ \frac{1}{\delta^2} -\frac{2}{\delta^3}+\frac{1}{\delta^4} \right].
\end{equation}
We  assumed  $d > 2 \rhoeclean$; for the general case just replace $2/\delta$ by $\vert \vert 2/\delta \vert \vert$. 

For cones  with maximum and minimum diameters $\rhoeclean \delta_{\rm M}$ and $\rhoeclean \delta_{\rm m}$:
\begin{equation}\label{LRVcone}
\LRVinicialde{cone}= \frac{16 \Omegaclean L }{3 \ln10} \left[\frac{3}{\Dm \delta_{\rm M}}-\frac{\delta_{\rm m}+\delta_{\rm M}}{(\delta_{\rm m} \delta_{\rm M})^2}+\frac{(\delta_{\rm m}+\delta_{\rm M})^2-\delta_{\rm m} \delta_{\rm M}}{(\delta_{\rm m} \delta_{\rm M})^3} \right].   
\end{equation}
Again we assumed $d(x)>2\rhoeclean$ for notational simplicity. This  result is valid both for increasing or decreasing cones, but note that  they will have different time-evolution after their equal initial filtration performance, see \sects{section3.2} and~\nosect{section3.3}.

For a sinusoidally corrugated tube of maximum and minimum diameters $\rhoeclean \delta_{\rm M}$ and $\rhoeclean \delta_{\rm m}$ with an integer number $n$ of oscillation periods:
\begin{equation} \label{LRVsinus}
\LRVinicialde{sinus} =  \frac{\Omegaclean  L }{\ln10}  \left[      \frac{8(\delta_{\rm M}+\delta_{\rm m})}{(\delta_{\rm M} \delta_{\rm m})^{3/2}} +\frac{4(3\delta_{\rm M}^2+2 \delta_{\rm M} \delta_{\rm m}+ 3\Dm^2)}{(\delta_{\rm M}\delta_{\rm m})^{5/2}}   +    \frac{(\delta_{\rm M}+\delta_{\rm m})(5\delta_{\rm M}^2-2\delta_{\rm M} \delta_{\rm m} + 5\delta_{\rm m}^2)}{(\delta_{\rm M}\delta_{\rm m})^{7/2}}            \right].
\end{equation}
Again we assumed $d(x)>2\rhoeclean$. The result is independent on $n$, as far as it is integer and nonzero.

The  LRV$(t)$   in     \fig{figuratres} (and \tabla{tablauno}) were obtained by numerical integration of \eqs{diferencialred} to\noeq{flujo}. Their limit values for $t\rightarrow0$  fully coincide with the results of \eqs{LRVcyl} to\noeq{LRVsinus}.


\begin{thebibliography}{99}
\bibitem{footnote-uno}We use the notation $\Vert x \Vert$ as a way to simplify our notation yet keeping our equations valid if the filter is beyond its saturation point. As the numerical evaluations  in this article are relevant only until that saturation point,  for most practical purposes $\Vert x \Vert=x$ in our present context.
\bibitem{footnote-omegacleanfinite}This is because the survival (\ie, non-binding) probabilities per unit length must accumulate multiplicatively (and not linearly) over successive differential lengths when considering a finite-length slice $(x_j,x_j+\Delta x)$. If we identify  the minimum length over which \Omegaclean\ makes sense with an impurity size $\rho_0$, that condition translates to $(1-\rho_0\Omegaclean)^{\Delta x /\rho_0}=1-\Omegacleanfinite\Delta x$ and hence to \eq{eq-omegacleanfinite}.
\bibitem{footnote-dos}It could be thought that the hydrodynamic energy  could be not the only one playing a role in this kind of filtration, and that the  binding energy per impurity should perhaps be also considered. However, the latter becomes negligible in our case, as may be seen by an order of magnitude comparison: The more energetic bond between  impurities and  walls would be an ionic one, of typically $10^{-18}$\,J per impurity. This is to be compared to  \eq{hydrod} or about $10^{-5}$\,J. Therefore,   for  the situation studied in this article  the total energy required for filtration may be identified with the hydrodynamic rate.
\end{thebibliography}
\end{document}